\documentclass[notitlepage,preprint]{article}
\usepackage[utf8]{inputenc} 
\usepackage[bitstream-charter]{mathdesign} 
\usepackage[colorlinks=true,urlcolor=blue,linkcolor=blue, citecolor=blue]{hyperref}
\usepackage{graphicx}
\usepackage[a4paper,left=2.5cm, right=2.5cm, top=2cm, bottom=2cm]{geometry}
\usepackage[noblocks]{authblk}

\usepackage{amsthm}
\usepackage{amsmath}

\title{LAGO: the Latin American Giant Observatory}

\author[1,*]{Iv\'an Sidelnik}
\author[2,3,**]{Hern\'an Asorey}
\author[4]{for the LAGO Collaboration}

\affil[1]{Departamento de Física de Neutrones y Reactores,
CONICET e Instituto Balseiro, Av. Bustillo 9500, (8400) S.C. de Bariloche,
Argentina.}
\affil[2]{Laboratorio Detecci\'on de Part\'{\i}culas y
Radiaci\'on, CNEA e Instituto Balseiro, Av. Bustillo 9500, (8400) S.C. de
Bariloche, Río Negro, Argentina.}
\affil[3]{Instituto de Tecnolog\'{\i}as en Detecci\'on y
Astropart\'{\i}culas (CNEA, CONICET, UNSAM), Av. Gral. Paz 1499, (1650) San
Martín, Buenos Aires, Argentina}
\affil[4]{The Latin American Giant
Observatory, \href{http://lagoproject.org}{lagoproject.org}, see the full list of members and institutions at \href{http://lagoproject.org/colab.html}{lagoproject.org/collab.html}}
\affil[*]{Corresponding author, \href{mailto:sidelnik@cnea.gov.ar}{sidelnik@cnea.gov.ar}}
\affil[**]{Corresponding author, \href{mailto:asoreyh@cab.cnea.gov.ar}{asoreyh@cab.cnea.gov.ar}}

\begin{document}

\maketitle

\begin{abstract}
	The Latin American Giant Observatory (LAGO) is an extended cosmic ray
	observatory composed of a network of water-Cherenkov detectors (WCD)
	spanning over different sites located at significantly different altitudes
	(from sea level up to more than $5000$\,m a.s.l.) and latitudes across
	Latin America, covering a wide range of geomagnetic rigidity cut-offs and
	atmospheric absorption/reaction levels. The LAGO WCD is simple and robust,
	and incorporates several integrated devices to allow time synchronization,
	autonomous operation, on board data analysis, as well as remote control and
	automated data transfer.

	This detection network is designed to make detailed measurements of the
	temporal evolution of the radiation flux coming from outer space at ground
	level.  LAGO is mainly oriented to perform basic research in three
	areas: high energy phenomena, space weather and atmospheric radiation at
	ground level.  It is an observatory designed, built and operated by the
	LAGO Collaboration, a non-centralized collaborative union of more than 30
	institutions from ten countries.

	In this paper we describe the scientific and academic goals of the LAGO
	project - illustrating its present status with some recent results - and
	outline its future perspectives.
\end{abstract}


\section{Introduction}


Astroparticle physics is nowadays one of the scientific fields that evidence
large interdisciplinary contributions. This is not only possible but even
needed, given the large array of topics this discipline covers. 
Several space-borne and ground-based cosmic ray observatories have been built
or are under development. These facilities are mainly coordinated in large
international collaborations with hundreds or even thousands of scientists. In
the particular case of Latin America (LA), the successful installation and
commissioning of the Pierre Auger
Observatory\,\cite{ThePierreAugerCollaboration2015} in Malarg\"ue, Argentina,
generated an outstanding opportunity to develop Astroparticle physics and High
Energy Physics in this region. In this work, we describe one of the resulting
projects of this regional development: the Latin American Giant Observatory
(LAGO). 

\section{The Latin American Giant Observatory}\label{sc_lago}

The Latin American Giant Observatory (LAGO), formerly known as the Large
Aperture Gamma Ray Bursts Observatory, is a project conceived in
2006\,\cite{Allard2008} to detect the high energy component of Gamma Ray Bursts
(GRBs, with typical energy of primaries $E_p \gtrsim 20$\,GeV) by installing
$10$\,m$^2-20$\,m$^2$ water Cherenkov detectors (WCD) at very high altitude
sites across the Andean ranges. From this initial aim, LAGO has evolved toward
an extended astroparticle observatory at a regional scale, currently operating
WCDs and other particle detectors at nine countries in LA (Argentina, Bolivia,
Brazil, Colombia, Ecuador, Guatemala, Mexico, Peru and Venezuela) together with
the recent incorporation of institutions from Spain. LAGO is operated by the
LAGO Collaboration, a cooperative and non-centralized collaboration of 26
institutions.

LAGO has three main scientific objectives: to study high energy gamma events at
high altitude sites, to understand space weather phenomena through monitoring
at the continental scale, and to decipher the impact (direct and indirect) of
the cosmic radiation on atmospheric phenomena. These objectives are
complemented by two main academic goals: to train students in astroparticle and
high energy physics techniques, and foremost, to support the development of
astroparticle physics in LA\,\cite{Asorey2015c}.

The LAGO detection network, consist of single or small arrays of astroparticle
detectors installed in different sites across the Andean
region\,\cite{Sidelnik2015}.  Currently, ten WCDs are in operation and we
expect to have another eleven detectors starting their operation and
calibration in the next biennium. Our detection network spans a region from
the south of Mexico, with a small array installed at Sierra Negra ($4550$\,m
a.s.l.), to Patagonia, with three WCDs installed at Bariloche ($865$\,m
a.s.l.). In addition, the installation of two WCDs was recently funded for the
Marambio Base (Arg., 200m a.s.l.) located in the Antarctic Peninsula, mainly
oriented for Space Weather studies and monitoring\,\cite{Asorey2015a}.

The network is distributed over similar geographical longitudes but a wide
range of geographical latitudes and altitudes. By combining simultaneous
measurements at different rigidity cutoffs from regions with differing
atmospheric absorption we are able to produce near-real-time information at
different energy ranges of, for example, disturbances induced by interplanetary
transients and long term space weather phenomena.

Each LAGO water Cherenkov detector consists of a plastic tank containing
$1$\,m$^3$ to $40$\,m$^3$ of purified water, where one to four top-mounted
large photomultiplier tubes (PMT) collect the Cherenkov light produced by
ultra-relativistic particles moving through the water volume. Typically, three
types of large photocatode area PMTs are used in LAGO WCDs: Photonis XP1805,
Hamamatsu R5912 and Electron Tubes 9353KB.  The water purification protocol
used includes physical (membranes filtration and sedimentation); chemical
(mainly coagulation and flocculation using alum, Al$_2$SO$_4$, and iron salts,
FeCl$_3$); and biological methods (activated carbon filtration and disinfectant
agents such as NaClO). With this method we achieve very low turbidity and high
purity water, with a mean resistivity of the order of $\sim$ M$\Omega$.  An
internal coating of an ultra-violet light highly reflective and diffusive
fabric is used. The WCD signals are shaped and digitized by a custom made
$40$\,MHz electronic board controlled by a Digilent Nexys2 FPGA.

Besides WCD signals, the need to measure very different physical magnitudes,
such as cloudiness, humidity and atmospheric pressure and temperature, in
remote places with almost no infrastructure requires autonomous and reliable
detectors and components. We have moreover based our design in commercial
off-the-shelf devices, which have to be available in all the countries where
LAGO operates\,\cite{Sidelnik2015}. The main component is the electronic
system that operates the complete station. Additional sensors are integrated in
an Arduino controlled common sensor module to measure atmospheric pressure, air
temperature, solar radiance, cloudiness, and even greenhouse and smog gas
concentrations. A commercial global navigation satellite system module is
included for timing synchronization and to allow ionospheric total electron
content (TEC) calculations at some sites\,\cite{Mannucci1998}. All these
devices, including the WCD electronics boards, are powered by solar panels and
are controlled by a single board computer that includes a GPRS module to
transfer telemetry and on board pre-analyzed data by using mobile phone
networking data services.  This new concept produces data that extends the
typical objectives of a cosmic ray observatory, allowing several scientific,
academic and even citizen science communities to take advantage of our regional
scale detection network\,\cite{Asorey2015c}. 

\section{The LAGO Programs}\label{sc_progs}

Scientific and academic objectives are organized in different programs and are
carried out by the corresponding working groups. LAGO programs cover several
aspects of the project, from the installation, calibration and operation of the
detectors to the search for pathways to transfer data from remote sites.
Complete simulation chains involving all the related aspects (from CR
propagation to detector response)\,\cite{Asorey2015a}, data analysis techniques
specially designed for the very different energy and temporal scales of the
studying phenomena\,\cite{Asorey2015a,Asorey2015b}, data preservation, and the
design of new experiments to be conducted in graduate and undergraduate
lectures in the participating Universities, are just some examples of the range
of the objectives of the LAGO project. 

High altitude sites ($h>4500$\,m) are designed and operated mainly for the
search for high energy components of GRB. Such sites are chosen to diminish the
atmospheric absorption of extensive air showers (EAS) initiated by low energy
cosmic rays. To increase electromagnetic-muon separation at single pulse level
in high-altitude sites, a method based on the total charge and pulse rise-time
analysis is also implemented. The increase in separation performance can be
used to improve the search for possible GRB candidates, as gamma initiated
showers show lower muon fractions at detector level when compared with hadronic
primaries. 

\begin{figure}[h] 
  \centering
   \includegraphics[width=0.50\columnwidth]{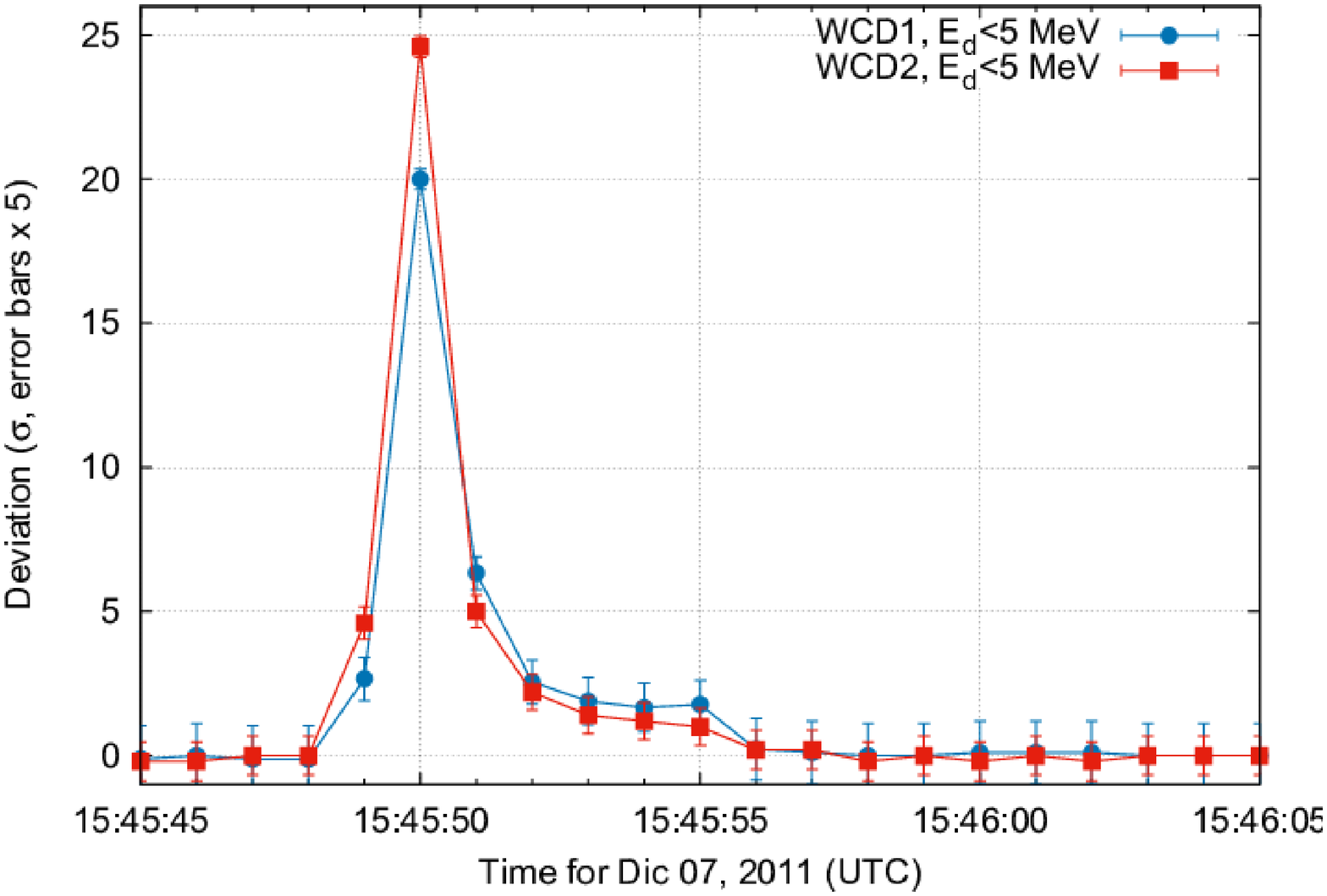}
   \includegraphics[width=0.49\columnwidth]{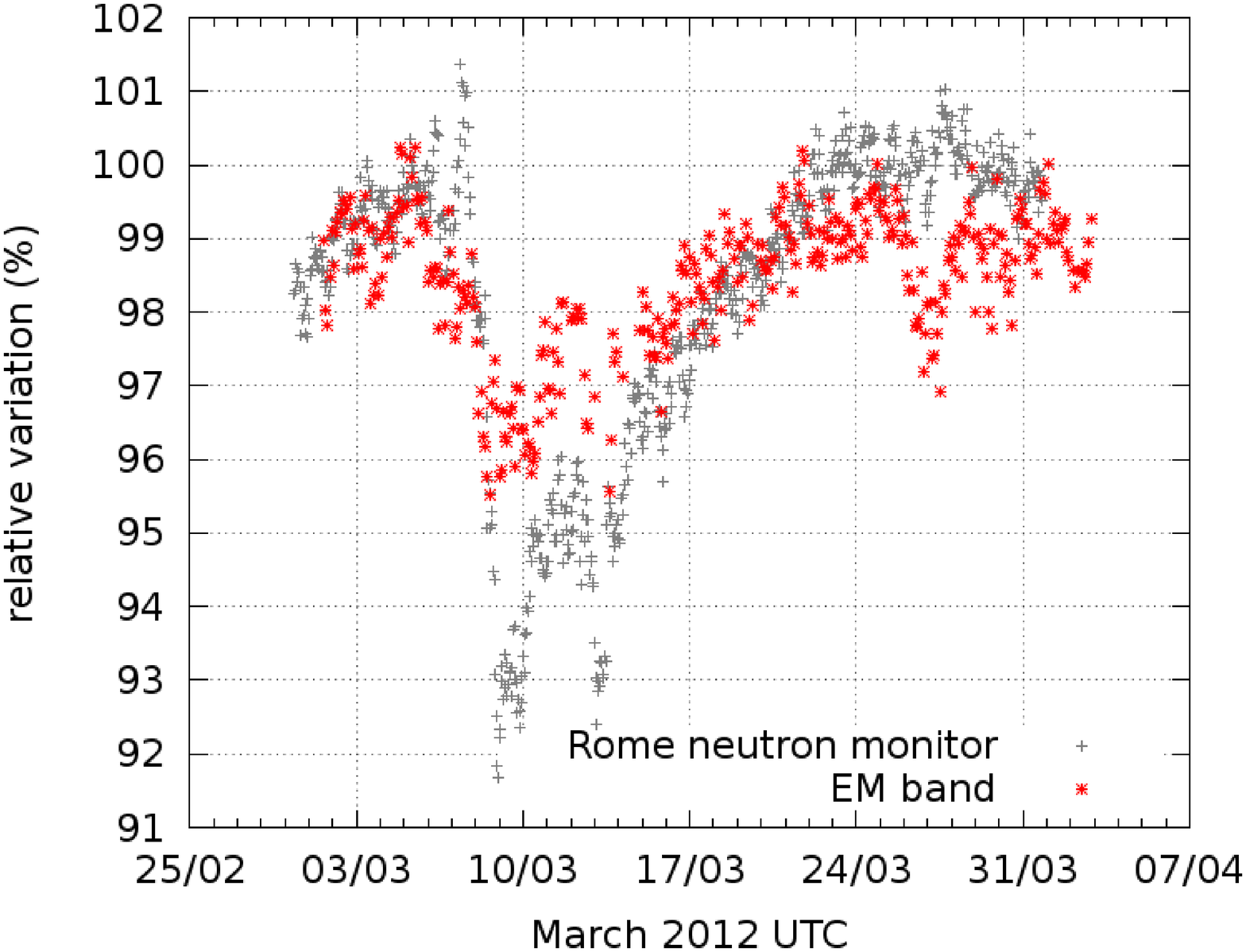}
	\caption {
		Left: Signal alert for a potential event candidate registered on Wed
		Dec 07 15:47:02.378 UTC 2011 detected by LAGO at Chacaltaya, Bolivia,
		5250 m a.s.l. (from\,\cite{Asorey2015b}).  Right: Multi-spectral
		analysis of the Forbush Decrease of March 8th, 2012, measured in a
		single $1.8$\,m$^2$ WCD in Bariloche, Argentina (red stars), compared
		with neutron flux measured by the Rome neutron monitor (gray pluses).
	}
	\label{fi_he}
\end{figure}

All the LAGO programs exploit the single particle technique (SPT) by looking
for significant excesses in the counting of background signals at different
sites. In particular, the LAGO facilities at Chacaltaya mountain consist of
three LAGO WCDs with a total detection area of 10\,m$^2$. In the period
2010-2012 the Chacaltaya station has collected data during $\sim 17000$\,hours
of detection for WCD1 and WCD2, and $\sim 15700$\,hours for WCD3. Recent
studies, based on CORSIKA simulations, show that the angular aperture of the
Chacaltaya site can be extended up to a zenith angle of $25^\mathrm{o}$ in the
energy range of interest for GRB and other Gamma originated signals. We
estimate that detection area should be ten times larger to extend sensitivity
up to $45^\mathrm{o}$. Combining these results with the uptime of each WCD in
this period, the total exposure of this site accumulated during this period was
$2.7 \times 10^8$\,m$^2$\,s\,sr. Considering the typical variations of the flux
at the time scales of interest, we produced a strict search criterion based on
an adaptive moving window averaging method (see \,\cite{Asorey2015b}).
Following this procedure, a new potential candidate was found, appearing on Wed
Dec 07 15:45:49.675$\pm$0.005 UTC 2011 and a duration $5.5$\,s, as can be seen
in the left panel of figure \ref{fi_he}. The equatorial
coordinates of Chacaltaya zenith were RA/Dec (J2000) $16^\mathrm{h}
17^\mathrm{m} 31.3^\mathrm{s} / -16^\circ 21' 00"$, with an acceptance aperture
of $\theta \lesssim 25^\circ$. At that time, the Fermi satellite was outside of
our acceptance cone in Chacaltaya.

While this technique has the lack of directional reconstruction, it is however
possible to use atmospheric absorption as a selection tool for periodic signals
including gamma point sources. Data stacking or summation in two different time
systems, solar and sidereal, were made over the data. The idea behind this
process is based on the random, Poissonian, nature of the majority of the
radiation measured by our detectors. When a non random and periodic signal
exists, data can be summed over repetitive occurrences, both in sidereal and
solar time, to increase the signal-to-noise ratio. After doing this, we
observed clear indications, both in phase and in amplitude, of solar modulation
of the flux of galactic cosmic rays. These observations demonstrate that, by
using an adapted analysis technique to the characteristics of our small
detectors, it is possible to observe space weather phenomena at different time
scales from ground level in the LAGO network of WCD.

Simultaneous measurements of galactic cosmic ray flux modulation at different
locations on Earth using the same type of detector can provide important
information on the properties and global structure of magnetic clouds reaching
the terrestrial environment during interplanetary Coronal Mass Ejections (ICME)
events pointing towards Earth\,\cite{Munakata2012}. By using WCDs, we are able
to determine the flux of secondary particles at different bands of deposited
energy within the detector volume.  These bands are dominated by different
components of the CR spectrum reaching the Earth's atmosphere (primaries). This
multi-spectral analysis technique (MSAT) constitutes the basis of our Space
Weather oriented data analysis\,\cite{Asorey2015a}. By combining all the data
measured at different locations of the detection network, the LAGO project will
provide very detailed and simultaneous information of the temporal evolution
and small- and large-scale characteristics of the disturbances produced by
different transient and long-term space weather phenomena.

A complex and complete chain of simulations support this
program\,\cite{Asorey2015a}. For every LAGO site, the
directional rigidity cutoff $R_c$ is determined for secular and altered
geomagnetic field conditions, such as those produced during intense geomagnetic
storms. The expected number of primaries is then determined by integrating the
measured flux of all the hadronic cosmic rays with $1<E_p/\mathrm{GeV}<10^6$. A
set of CORSIKA simulations is then used to determine the expected number of
secondary particles at ground level.  Only those secondaries originating from
primaries with rigidities above local cut-offs are included. Finally, the
expected secondary particles are injected into a Geant4-based model of the LAGO
WCD to obtain the signal light flux at the detector level.

With the LAGO MSAT we are able to determine the evolution of the flux observed
in three different bands. As the transition points between these different
regimes are characterized by changes in the histogram slopes, a fully automated
algorithm searches for all these features in 1-hour calibration histograms. As
an example, in the right panel of figure \ref{fi_he} the measurement of the
Forbush decrease occurred on March 8th, 2012, is shown for the electromagnetic
dominated band. These measurements demonstrate the capabilities of WCDs to
extend present studies of space weather phenomena using low cost detectors from
ground level.

\section{Conclusions}\label{sc_concs}

In this paper we outline the Latin American Giant Observatory, its main
scientific capabilities and implications for the development of astroparticle
physics at Latin America. We also describe the key programs of LAGO, oriented
to search for GRBs, high energy phenomena, and space weather, using low cost
and reliable detectors at ground level. A complete chain of simulations and
data analysis techniques has been developed to exploit WCD measuring
capabilities in different energy regimes. Several cited examples demonstrate
the LAGO capabilities to study very different astrophysical phenomena.  With
new detectors coming into operation (for example at as the new LAGO site in
Antarctica) a large and unprecedented detection network is emerging at a
regional scale.

Besides its scientific objectives, the LAGO project is also a seeder for
astroparticle physics development in the LA. A proof of this is the
fact that most of the results described in this work have been obtained by
several undergraduate and master students working in coordinated collaboration
in several Latin American countries. LAGO is a promising observatory that is
helping to support the development of several interdisciplinary branches
associated with Astroparticles in Latin America.

\paragraph*{Acknowledgments}

The LAGO Collaboration is very thankful to all the
participating institutions and to the Pierre Auger Collaboration for their
continuous support. 



\begin{thebibliography}{1}
\expandafter\ifx\csname url\endcsname\relax
  \def\url#1{\texttt{#1}}\fi
\expandafter\ifx\csname urlprefix\endcsname\relax\def\urlprefix{URL }\fi
\expandafter\ifx\csname href\endcsname\relax
  \def\href#1#2{#2} \def\path#1{#1}\fi

\bibitem{ThePierreAugerCollaboration2015}
{The Pierre Auger Collaboration}, {The Pierre Auger Cosmic Ray Observatory}
  (2015) 1--97\href {http://dx.doi.org/10.1016/j.nima.2015.06.058}
  {\path{doi:10.1016/j.nima.2015.06.058}}.

\bibitem{Allard2008}
D.~Allard, I.~Allekotte, C.~Alvarez, et~al., {Use of water-Cherenkov detectors
  to detect Gamma Ray Bursts at the Large Aperture GRB Observatory (LAGO)},
  Nuclear Instruments and Methods in Physics Research Section A 595~(1) (2008)
  70--72.
\newblock \href {http://dx.doi.org/10.1016/j.nima.2008.07.041}
  {\path{doi:10.1016/j.nima.2008.07.041}}.

\bibitem{Asorey2015c}
H.~Asorey, {for the LAGO Collaboration}, {LAGO: the Latin American Giant
  Observatory}, in: Proceedings of Science, The Hague, 2015, p.
  PoS(ICRC2015)247.

\bibitem{Sidelnik2015}
I.~Sidelnik, {for the LAGO Collaboration}, {The Sites of the Latin American
  Giant Observatory}, in: Proceedings of Science, The Hague, 2015, p.
  PoS(ICRC2015)665.

\bibitem{Asorey2015a}
H.~Asorey, S.~Dasso, L.~A. N{\'{u}}{\~{n}}ez, et~al., {The LAGO Space Weather
  Program: Directional Geomagnetic Effects, Background Fluence Calculations and
  Multi-Spectral Data Analysis}, Proceedings of Science ICRC2015 (2015)
  PoS(ICRC2015)142.

\bibitem{Mannucci1998}
A.~J. Mannucci, B.~D. Wilson, et~al., {A global mapping technique for
  GPS-derived ionospheric total electron content measurements}, Radio Science
  33~(3) (1998) 565--582.
\newblock \href {http://dx.doi.org/10.1029/97RS02707}
  {\path{doi:10.1029/97RS02707}}.

\bibitem{Asorey2015b}
H.~Asorey, P.~Miranda, A.~N{\'{u}}{\~{n}}ez-Casti{\~{n}}eyra, L.~A.
  N{\'{u}}{\~{n}}ez, J.~Salinas, C.~{Sarmiento Cano}, R.~Ticona, A.~Velarde,
  {Analysis of Background Cosmic Ray Rate in the 2010-2012 Period from the
  LAGO-Chacaltaya Detectors}, in: Proceedings of Science, The Hague, 2015, p.
  PoS(ICRC2015)414.

\bibitem{Munakata2012}
K.~Munakata, {Probing the heliosphere with the directional anisotropy of
  galactic cosmic-ray intensity}, Proceedings of the IAU 7~(S286) (2012)
  185--194.
\newblock \href {http://dx.doi.org/10.1017/S1743921312004826}
  {\path{doi:10.1017/S1743921312004826}}.

\end{thebibliography}

\end{document}